\documentclass[sigconf]{acmart}

\usepackage{siunitx}        
\usepackage{subcaption}           
\usepackage{algorithm}
\usepackage{algpseudocode}
\usepackage[nameinlink,capitalize]{cleveref} 

\usepackage{tikz}
\usetikzlibrary{shapes.geometric}
\tikzset{
  shapeStyle/.style={draw, thick, fill=gray!20},
  point/.style={draw, red, fill=red, circle, inner sep=1pt}
}

\setcopyright{acmlicensed}
\copyrightyear{2026}
\acmYear{2026}
\acmDOI{XXXXXXX.XXXXXXX} 
\acmConference[SIGSPATIAL '26]{The 34th ACM SIGSPATIAL International Conference on Advances in Geographic Information Systems}{November 3--6, 2026}{Riverside, CA, USA}
\acmBooktitle{The 34th ACM SIGSPATIAL International Conference on Advances in Geographic Information Systems (SIGSPATIAL '26), November 3--6, 2026, Riverside, CA, USA}
\acmISBN{979-8-4007-XXXX-X/26/11}

\begin{document}

\title{Rapid Quantification of Outdoor Object Visibility in Urban Setting Using Connected-Vehicle Fields of View}

\author{Artur Grigorev}
\correspondingauthor
\orcid{0000-0001-6875-3568}
\email{artur.grigorev@uts.edu.au}
\affiliation{%
  \institution{Faculty of Engineering and IT, University of Technology Sydney}
  \city{Sydney}
  \country{Australia}}

\author{Adriana-Simona Mihăiţă}
\orcid{0000-0001-7670-5777}
\email{adriana-simona.mihaita@uts.edu.au}
\affiliation{%
  \institution{Faculty of Engineering and IT, University of Technology Sydney}
  \city{Sydney}
  \country{Australia}}

\renewcommand{\shortauthors}{Grigorev and Mih\u{a}i\c{t}\u{a}}

\begin{abstract}
Identifying locations that offer maximum visual exposure to passing vehicular traffic is a core problem in urban analytics, with applications spanning urban design, navigation, location-based services, and the placement of street-level assets. Traditional site selection methods often rely on static traffic counts or subjective assessments. This research introduces a data-driven methodology to objectively quantify location visibility by analyzing large-scale connected vehicle trajectory data within urban environments. We model the dynamic driver field-of-view using a forward-projected visibility area for each vehicle position derived from interpolated trajectories. By integrating this with building vertex locations extracted from OpenStreetMap, we quantify the cumulative visual exposure, or ``visibility count'', for thousands of potential points of interest along roadways. The core technical contribution involves the construction of a BallTree spatial index over building vertices. This enables highly efficient (O(logN) complexity) radius queries to determine which vertices fall within the viewing circles of millions of trajectory points across numerous trips, significantly outperforming brute-force geometric checks. Analysis reveals two key findings: 1) Visibility is highly concentrated, identifying distinct 'visual hotspots' receiving disproportionately high exposure compared to average locations. 2) The aggregated visibility counts across vertices conform to a Log-Normal distribution.
\end{abstract}



\maketitle

\section{Introduction}

Understanding which features of the built environment are actually seen by people moving through a city is a long-standing problem with applications across urban design, navigation, location-based analytics, and the planning of street-level infrastructure such as signage and Out-of-Home (OOH) advertising assets \cite{Madlenak2023}. Traditional approaches to quantifying such visual exposure often rely on static demographic data or aggregated traffic volumes, which may inadequately capture the dynamic reality of urban mobility and actual visibility \cite{Clow2023Integrated}. Simple Opportunity To See (OTS) metrics often overestimate true exposure, failing to account for viewing angles, vehicle dynamics (e.g. duration of stops), obstructions, and travel paths \cite{Madlenak2023}.

To address these limitations, data-driven approaches that take advantage of real-world movement data are gaining traction. Connected vehicle technologies generate vast amounts of granular trajectory data, offering a detailed view of how vehicles navigate the urban landscape. Concurrently, open geospatial datasets provide information about the geometry of built environment. This research focuses on utilizing these data sources to develop a more precise, observation-based understanding of visibility, specifically targeting the exposure of urban building features from the perspective of vehicular traffic.

This paper introduces a methodology to quantify the visibility of urban building vertices using connected vehicle trajectory data (sourced from the Compass IoT platform, an Australian startup collecting trajectory and telemetry data from connected vehicles across Australia, New Zealand, UK and USA) and building footprint data (from OpenStreetMap). We address the computational challenge of analysing large-scale trajectory data against complex urban geometry through efficient spatial indexing techniques, specifically BallTree structures. Our approach involves: (1) processing and interpolating connected vehicle trajectories to regular time intervals and calculating instantaneous vehicle bearing; (2) defining a dynamic, forward-looking viewing area for each vehicle position; (3) extracting vertices from OpenStreetMap building polygons; (4) utilizing a BallTree index for efficient querying of building vertices within the vehicles' viewing areas; and (5) aggregating visibility counts per building vertex across numerous trajectories.

The primary contribution of this work is a scalable framework for generating granular, objective metrics of building vertex visibility based on observed traffic patterns. This provides a quantitative basis for identifying high-exposure locations (``visual hotspots'') relevant to a range of applications, including informing urban design, enhancing navigation systems, supporting location-based analytics, and guiding the placement of street-level assets such as signage, public information displays, and other visibility-sensitive infrastructure.

\section{Related Works}
\label{sec:related_works}

The challenge of quantifying how urban features are seen from the road, and of using such estimates for downstream tasks such as siting visible infrastructure or assessing advertising exposure, has spurred research into data-driven methodologies that move beyond traditional static analyses \cite{Madlenak2023}. Traditional methods, often relying on aggregated traffic counts or area demographics \cite{Clow2023Integrated}, usually do not capture the dynamic nature of driver mobility and the nuances of actual visibility \cite{Madlenak2023}. This review situates our work within the context of two primary data-driven approaches identified in the literature: traffic simulation and real-world trajectory data analysis.

Traffic simulation including macroscopic, microscopic, and agent-based models (ABM), is a well-established tool in transportation research for analyzing network performance, safety, and demand \cite{liu2024modeling, DevelopmentEvaluationSimulation, AnAgentBasedModel}. While microscopic and ABM simulations offer the potential to model individual viewer (pedestrian or vehicle) paths and incorporate environmental factors affecting visibility \cite{liu2024modeling, AgentbasedSimulationPedestrian}, their documented application specifically for quantifying geometric OOH visibility or optimizing placement based on such metrics appears limited. Approaches like Cellular Automata (CA) have been used to model the reach of mobile transit advertising, focusing on information spread rather than geometric shape of visibility \cite{ModellingImpactTransit}.

Our research aligns more closely with approaches utilizing real-world trajectory data. This data, sourced from GPS devices, mobile phones, or public transport smart cards, captures observed movement patterns \cite{HandbookMobilityDataMining}. Analysis techniques applied to such data include preprocessing \cite{StudyMapMatching}, clustering common routes \cite{ScalableFrameworkTrajectoryPrediction}, predicting destinations \cite{SolvingDataSparsity}, and inferring trip purpose \cite{TargetedAdvertisingThesis}, often with the goal of enabling targeted advertising. For instance, research using South East Queensland's smart card data demonstrated optimizing ad placement in transit networks based on inferred passenger activities \cite{TargetedAdvertisingThesis}. Similarly, destination prediction algorithms, developed partly by Australian researchers using taxi GPS data, aim to inform relevant advertising along predicted routes \cite{SolvingDataSparsity}. The work by Fong et al. (2013) \cite{Fong2013Identifying} explores optimal spatial groups in sensor networks to maximize coverage or effect using clustering and optimization techniques. The obtained visibility data in our research could potentially inform optimal roadside sensor placements, identifying locations frequently observed by vehicular traffic.

A particularly relevant example is the SmartAdP visual analytics system \cite{SmartAdP}. SmartAdP uses large-scale historical taxi trajectory data to help planners select optimal billboard locations. It derives metrics like traffic volume, speed, reach (coverage), and Opportunities-To-See (OTS) directly from the trajectory data and relies on spatial indexing for efficient querying \cite{SmartAdP}. While highly relevant in its use of trajectory data and focus on placement optimization, SmartAdP differs from our approach in key ways: it relies on historical taxi data (which may not represent all traffic), uses derived metrics as proxies for exposure rather than direct geometric visibility calculation, and focuses on billboard locations rather than granular building vertex visibility.

Furthermore, the concept of visibility itself is complex. Industry standards like Australia's MOVE system attempt to quantify it by adjusting OTS based on factors like viewing angle, speed, distance, and ad characteristics, using data from travel surveys and eye-tracking rather than dynamic simulation or granular trajectory-based geometric checks \cite{Madlenak2023, xiao2022generalized}. Other research domains utilize Geographic Information Systems (GIS) with line-of-sight (LoS) or viewshed analyses \cite{KEROUANTON2024100734, gschwend2015relating, Chmielewski2017}, typically for static viewpoints or environmental assessment, highlighting the geometric tools potentially adaptable for dynamic analysis.

Our work contributes to the area by: 1) Utilizing high-frequency (up to every 5 second trajectory updates) connected vehicle location and direction data which may offer broader coverage of traffic (includes passenger vehicles, buses,  than specific sources like taxis or smart cards. 2) Focusing specifically on quantifying the geometric visibility of discrete urban features (building vertices from OSM) rather than relying solely on proximity or traffic volume metrics. 3) We also use BallTree spatial indexing explicitly for the efficient execution of numerous dynamic point-in-circle visibility queries derived from vehicle trajectories, enabling scalability. By combining these elements, we aim to provide a granular and computationally efficient method for assessing urban visibility potential from the perspective of real moving vehicles.

\subsection{Study Area}
\label{sec:study_area}

The geographical focus of this research is the inner-city suburb of \textbf{Waterloo} and its immediate surroundings, located approximately 3-4 kilometers south of the Sydney Central Business District (CBD) in New South Wales, Australia. Waterloo is characterized by a dense urban landscape, featuring a mix of historical residential terraces, contemporary high-density apartment complexes, commercial buildings, cafes and public parks.

\begin{figure}[h!]
\centering

\includegraphics[width=\linewidth]{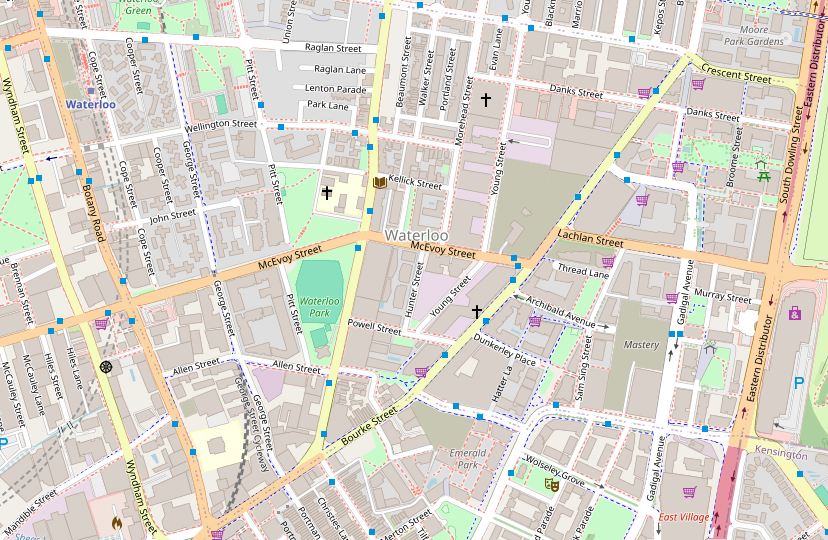}
\Description{An OpenStreetMap view of Waterloo, Sydney: a dense irregular street grid dividing the area into grey building blocks, with a few green park patches.}

\caption{Map of the Waterloo study area. (Image source: OpenStreetMap)}

\label{fig:waterloo_map}

\end{figure}

The specific region subjected to data analysis in this study is defined by the following geographical bounding box, using the \textbf{WGS84 coordinate system (EPSG:4326)}:
\begin{table}[h!] 
\centering
\caption{Geographical Bounding Box (WGS84 Coordinate System - EPSG:4326)}
\label{tab:bounding_box}
\begin{tabular}{l c c}
\hline
\textbf{Coordinate Type} & \textbf{Minimum Value} & \textbf{Maximum Value} \\
\hline
Longitude & $151.18943^{\circ}$ E & $151.21681^{\circ}$ E \\
Latitude  & $-33.91441^{\circ}$ S & $-33.89325^{\circ}$ S \\
\hline
\end{tabular}
\end{table}

This bounding box encapsulates the core of Waterloo and includes portions of adjacent suburbs such as Redfern, Zetland, and Alexandria. This selection provides a representative sample of inner-Sydney urban morphology, featuring a variety of road network types (major arterials, local streets) and mix of building densities, suitable for analyzing vehicular visibility patterns in a complex built environment. Building footprint data from OpenStreetMap and connected vehicle trajectory data from Compass IoT (limited to 2000 trips over 1 week of observations from the 1-7 march, year 2024) were filtered and analyzed specifically within this defined area.

\section{Methodology}
\label{sec:methodology}

The foundation for this analysis comprises two key processed datasets derived from raw sources. Firstly, we utilize a set of pre-processed connected vehicle trajectories, sourced from Compass IoT, which have been interpolated to provide a consistent vehicle position (latitude, longitude) and calculated bearing every 5 seconds within the Waterloo study area. Secondly, building footprint data obtained from OpenStreetMap. Instead of relying solely on the original vertices defining the building polygons, we interpolate additional points along each polygon edge segment. This ensures that the representation of the building outline consists of points spaced no more than 10 meters apart. This densified set of building points serves as the target features for our visibility assessment. The core objective of the subsequent methodology is to determine how frequently these densified building points fall within the dynamic viewing area projected forward from the moving vehicles.

Urban visibility assessment demands a geometric framework that balances analytical precision with computational scalability for large-scale transportation analytics. While vehicle forward-facing fields of view (FoVs) exhibit inherent angular constraints - typically 60-120 degrees in real-world driving scenarios - the primary objective here is not to model physical sensor limitations but to establish a computationally efficient baseline metric for visibility assessment.

This work adopts the \textbf{circle} as the foundational geometric approximation of driver FoV, explicitly prioritizing computational efficiency over precision.  Processing $O(10^6)$ vehicle trajectories with $O(10^3)$ building points requires sub-second per-point operations to enable feasible large-scale analysis. The circle containment test ($O(1)$) executes in 6 arithmetic operations (2 subtractions, 2 multiplications, 1 addition, 1 comparison), while sector-based alternatives require $\sim$100-250 operations due to the $\arctan$ function (Table~\ref{tab:shape-algorithms}).

\begin{table}[H]
\small
\caption{Geometric containment algorithms and their computational cost}
\label{tab:shape-algorithms}
\begin{tabular}{|p{0.20\linewidth}|p{0.35\linewidth}|p{0.3\linewidth}|}

\hline
\textbf{Shape \& Algorithm} & \textbf{Formula / Concept} & \textbf{Computational Cost (Operations)} \\
\hline \hline

\textbf{Circle}
\newline (Squared Distance Comparison)
\newline
\begin{tikzpicture}
    \node[shapeStyle, circle, minimum size=1.5cm] (c) at (0,0) {};
    \node[point] at (-0.2,0.2) {};
\end{tikzpicture}
& Compares the squared distance from point $P(p_x, p_y)$ to center $C(c_x, c_y)$ with the squared radius $r^2$. This avoids a slow square root operation.
\newline\newline
$(p_x - c_x)^2 + (p_y - c_y)^2 \leq r^2$
& \textbf{Fastest:}
\newline 2 Subtractions
\newline 2 Multiplications
\newline 1 Addition
\newline 1 Comparison
\newline \textbf{(Total: 6)} \\
\hline

\textbf{Triangle}
\newline (Cross Product Test)
\newline
\begin{tikzpicture}
    \node[shapeStyle, regular polygon, regular polygon sides=3, minimum size=1.5cm] (t) at (0,0) {};
    \node[point] at (0,0) {};
\end{tikzpicture}
& Checks if point $P$ is on the same side of all three edges (e.g., AB, BC, CA) by verifying the sign of the 2D cross product.
\newline\newline
For edge AB:
$z = (B_x - A_x)(P_y - A_y) - (B_y - A_y)(P_x - A_x)$
\newline
(Signs of all 3 results must be consistent)
& \textbf{Fast:}
\newline 9 Subtractions
\newline 6 Multiplications
\newline 3 Comparisons
\newline \textbf{(Total: 18)} \\
\hline

\textbf{Sector}
\newline (Composite Check)
\newline
\begin{tikzpicture}
    \path[shapeStyle] (0,0) -- (45:1cm) arc (45:135:1cm) -- cycle;
    \node[point] at (90:0.5cm) {};
\end{tikzpicture}
& A two-part test:
\newline 1. Point must be inside the circle.
\newline 2. The angle of the point must be within the sector's angular bounds.
\newline\newline
$(p_x-c_x)^2+(p_y-c_y)^2 \leq r^2 \quad \land$
\newline
$\theta_{start} \leq \text{atan2}(p_y-c_y, p_x-c_x) \leq \theta_{end}$
& \textbf{Slow:}
\newline 4 Subtractions
\newline 2 Multiplications
\newline 1 Addition
\newline 3 Comparisons
\newline A computationally expensive \textbf{`atan2`} function (depends on the implementation). \\
\hline

\textbf{Polygon ($n$-sided)}
\newline (Ray Casting Algorithm)
\newline
\begin{tikzpicture}
    \node[shapeStyle, regular polygon, regular polygon sides=6, minimum size=1.7cm] (p) at (0,0) {};
    \node[point] at (0.3, -0.2) {};
\end{tikzpicture}
& A Ray Casting Method counts how many times a ray extending from point $P$ intersects the polygon's edges. An odd count means the point is inside.
\newline\newline
$\text{is\_inside} = (\text{intersection\_count} \pmod 2 = 1)$
\newline
(Requires a loop over all $n$ edges)
& \textbf{Slowest:}
\newline A loop of $n$ iterations. An optimized test may avoid slow division operation. The worst-case cost per iteration is:
\newline 4 Subtractions
\newline 2 Multiplications
\newline 5 Comparisons
\newline \textbf{(Total: $\approx$ 11 x n)} \\
\hline
\end{tabular}
\end{table}

\subsection*{Selection of an Optimal Circular Sensor Approximation}

For computational efficiency in real-time applications, it is often advantageous to approximate complex sensor geometries, such as a sector, with simpler shapes like a circle. A quantitative evaluation was performed to identify the optimal circular approximation for a sensor with a 100-meter range and a 60-degree field of view. Several candidate circles, defined by their maximum extent along the x-axis ($x_{\text{end}}$), were analyzed using a Monte Carlo simulation.

To provide a metric of similarity between the true sector (Area $A$) and each circular approximation (Area $B$), we utilize the Sørensen-Dice coefficient ($D$). This metric balances the penalties for both under-approximating (missed area) and over-approximating (excess area). It is defined as:
\begin{equation}
    D = \frac{2 |A \cap B|}{|A| + |B|}
\end{equation}
A Dice coefficient of 1 indicates a perfect match, while 0 indicates no overlap. The results of this analysis, including the percentage of the sector's area covered (Overlap), missed, and incorrectly included (Excess), are detailed in Table~\ref{tab:circle_approx_eval}.

\begin{table}[htbp]
\centering
\caption{Evaluation of circular approximations for a 60 degree visibility sector. The table shows the percentage of the sector's area that is correctly covered (Overlap), missed (Missed), and incorrectly included (Excess). The Dice Coefficient measures the overall similarity, with the highest value indicating the best geometric fit.}
\label{tab:circle_approx_eval}
\begin{tabular}{lrrrr}
\toprule
$x_{\text{end}}$ (m) & Overlap \% & Missed \% & Excess \% & Dice Coeff \\
\midrule
80.00 & 58.63 & 41.47 & 37.51 & 0.60 \\
90.00 & 74.15 & 25.95 & 47.48 & 0.67 \\
95.00 & 82.57 & 17.52 & 52.88 & 0.70 \\
\textbf{100.00} & 91.47 & 8.63 & 58.57 & \textbf{0.73} \\
105.00 & 96.78 & 3.32 & 68.57 & 0.73 \\
110.00 & 99.28 & 0.82 & 82.12 & 0.71 \\
115.47 & 100.10 & 0.00 & 99.71 & 0.67 \\
\bottomrule
\end{tabular}
\end{table}

\begin{figure*}[htbp]
    \centering
    \includegraphics[width=0.5\textwidth]{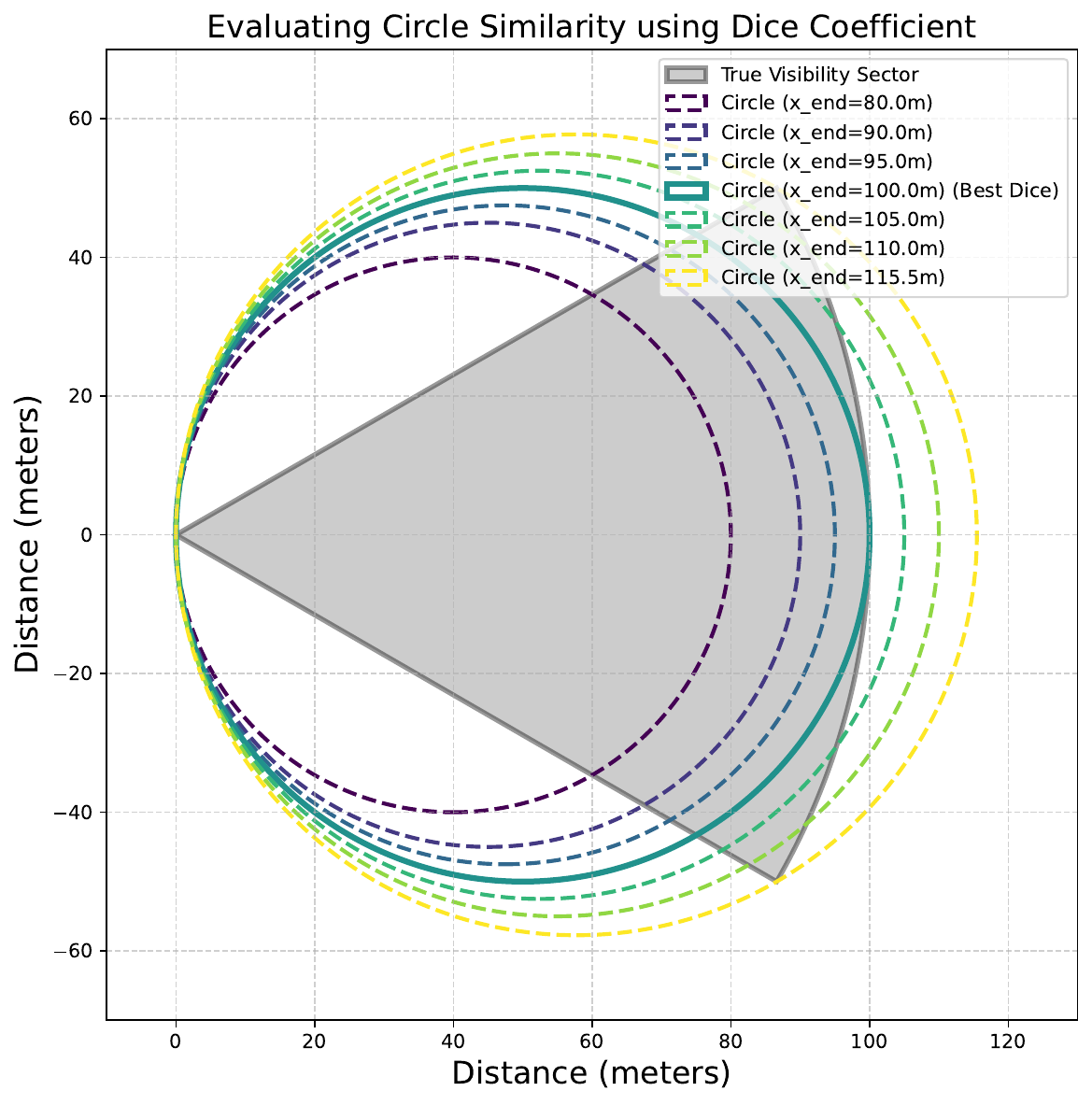}
    \Description{A plot in metres with a grey filled 60-degree sector opening rightward from the origin. Eight dashed circles tangent at the origin (labelled 80 to 115.5 metres) span a yellow-to-purple scale; the solid teal 100-metre circle, marked best Dice, fills the sector most closely.}
    \caption{Visual comparison of circular approximations to the true sensor sector. The circle with $x_{\text{end}}=100.0$~m, highlighted with a solid line, provides the best overall similarity as measured by the Dice Coefficient}
    \label{fig:circle_approx_plot}
\end{figure*}

We observe a clear trade-off: increasing the circle's size reduces the \textit{Missed \%} at the cost of increasing the \textit{Excess \%}. While the circle with $x_{\text{end}} = 115.47$~m guarantees complete coverage, which is critical for safety-first applications, it does so with substantial over-approximation, leading to a suboptimal Dice coefficient of 0.67. The approximation with $x_{\text{end}} = 100.00$~m achieves the maximum Dice coefficient of 0.73. This indicates that it provides the best overall coverage, representing the most effective balance between capturing the true sensor area (91.5\% overlap) and minimizing extraneous coverage. Consequently, for a balanced representation, the circle defined by $x_{\text{end}} = 100.00$~m is selected as the optimal approximation. The visual comparison of these approximations is shown in Figure~\ref{fig:circle_approx_plot}.

\subsection{Spatial Querying and Visibility Calculation}
\label{sec:querying}

Addressing the computational intensity of determining which densified building points fall inside numerous viewing circles requires an efficient spatial querying strategy.

\textbf{BallTree Spatial Indexing:} The core technique used is spatial indexing using a BallTree structure. This approach avoids brute-force distance calculations for every building point-circle pair, offering significant performance gains. The construction process involves several key steps: 1) Latitude/longitude coordinates of the densified building edge points (generated as described previously) are converted to radians, suitable for spherical calculations. 2) The BallTree index is constructed using these radian coordinates, specifically utilizing the ``haversine'' metric to ensure accurate great-circle distance computations. 3) A mapping is maintained to link the internal index of each point within the BallTree structure back to its unique identifier (representing its position along a specific building edge). The BallTree hierarchically partitions the building point data into nested hyperspheres (balls), enabling rapid pruning of the search space during distance-based queries.

\textbf{Point-in-Circle Queries:} With the BallTree index in place, the visibility calculation proceeds for each interpolated trajectory point possessing a valid bearing. The querying process follows these steps: 1) The viewing circle center is calculated (50m ahead along the bearing) using the \texttt{CalculateDestinationPoint} function. 2) The physical radius (50 meters) is converted into an equivalent angular radius ($R_{angular}$) based on the Earth's radius, appropriate for the Haversine distance metric used by the BallTree. 3) The BallTree's \texttt{query\_radius} method is invoked, supplying the circle center coordinates (in radians) and the calculated angular radius. This efficiently retrieves the indices of all densified building points located within the specified 50-meter great-circle distance from the viewing circle's center. 4) For each building point index returned by the query, a corresponding counter (representing the visibility count for that specific point during the trip) is incremented. This procedure is repeated for every valid point across all trajectories processed.

The detailed procedure for calculating these visibility counts for a single trip, operating on the densified building points, is formalized in Algorithm~\ref{alg:visibility_calc_points}.

\begin{algorithm*}[htbp]
\caption{Visibility Calculation for a Single Trip using BallTree (Densified Building Points)}
\label{alg:visibility_calc_points}
\begin{algorithmic}[1]
\Require Trip data $T$ (list of points with lat, lon, bearing)
\Require Densified building point spatial index $B_{tree}$ (BallTree)
\Require Mapping $M_{id}$ from tree index to original point identifier $p_{id}$
\Require Circle radius $R_{meters}$
\Ensure Map $V_{trip}$ of \{point\_id: visibility\_count\} for trip $T$

\State $N_{points} \gets \Call{GetNumberOfPoints}{B_{tree}}$ \Comment{Total number of densified points}
\State Initialize integer array $C_{point\_counts}$ of size $N_{points}$ with zeros
\State $R_{angular} \gets R_{meters} / \text{Earth\_Radius\_Meters}$ \Comment{Convert radius to radians}

\For{each point $p$ in $T$}
    \If{$p.bearing$ is valid}
        \State $(lat_c, lon_c) \gets \Call{CalculateDestinationPoint}{p.lat, p.lon, p.bearing, R_{meters}}$ \Comment{Calculate circle center}
        \If{$lat_c$ is valid \textbf{and} $lon_c$ is valid}
            \State $coord\_c^{rad} \gets \Call{ToRadians}{[[lat_c, lon_c]]}$ \Comment{Convert center to radians}
            \State $indices \gets \Call{QueryBallTreeRadius}{B_{tree}, coord\_c^{rad}, R_{angular}}$ \Comment{Find points in circle}
            \For{each index $idx$ in $indices[0]$} \Comment{Query returns list of lists}
                \State $C_{point\_counts}[idx] \gets C_{point\_counts}[idx] + 1$
            \EndFor
        \EndIf
    \EndIf
\EndFor

\State Initialize empty map $V_{trip}$
\For{$i \gets 0$ to $N_{points} - 1$}
    \If{$C_{point\_counts}[i] > 0$}
        \State $p_{id} \gets M_{id}[i]$ \Comment{Get original point ID from map}
        \State $V_{trip}\{p_{id}\} \gets C_{point\_counts}[i]$ \Comment{Store count in result map}
    \EndIf
\EndFor

\State \Return $V_{trip}$
\end{algorithmic}
\end{algorithm*}

\subsection{Aggregation and Analysis}
\label{sec:aggregation_analysis}

Following the per-trip visibility calculations which yield counts for densified building points, the results are consolidated to provide an overall measure of visibility for each point across the entire dataset.

\subsubsection*{Visibility Count Aggregation}
The primary goal here is to aggregate the visibility counts obtained for each densified building point during individual trips into a single total count. The aggregation logic proceeds as follows: 1) Per-trip visibility results (mapping point identifiers to counts) are processed, grouping counts by unique densified building point. A key derived from rounded coordinates is typically used to ensure these points are uniquely identified across different processing batches or potential floating-point discrepancies. 2) For each unique building point key, the visibility counts accumulated across all trips are summed to yield a final total visibility count. 3) The aggregated results are stored, in a structure mapping the unique point key back to its original coordinates and the calculated total visibility count. This aggregated count represents the overall frequency with which each densified building point along building outlines was found within a vehicle's viewing area across the studied trajectories.

\subsection{Trajectory and Visibility Visualization}
\label{sec:trajectory_visualization} 

\begin{figure*}[htbp]
    \centering
    \includegraphics[width=0.7\linewidth]{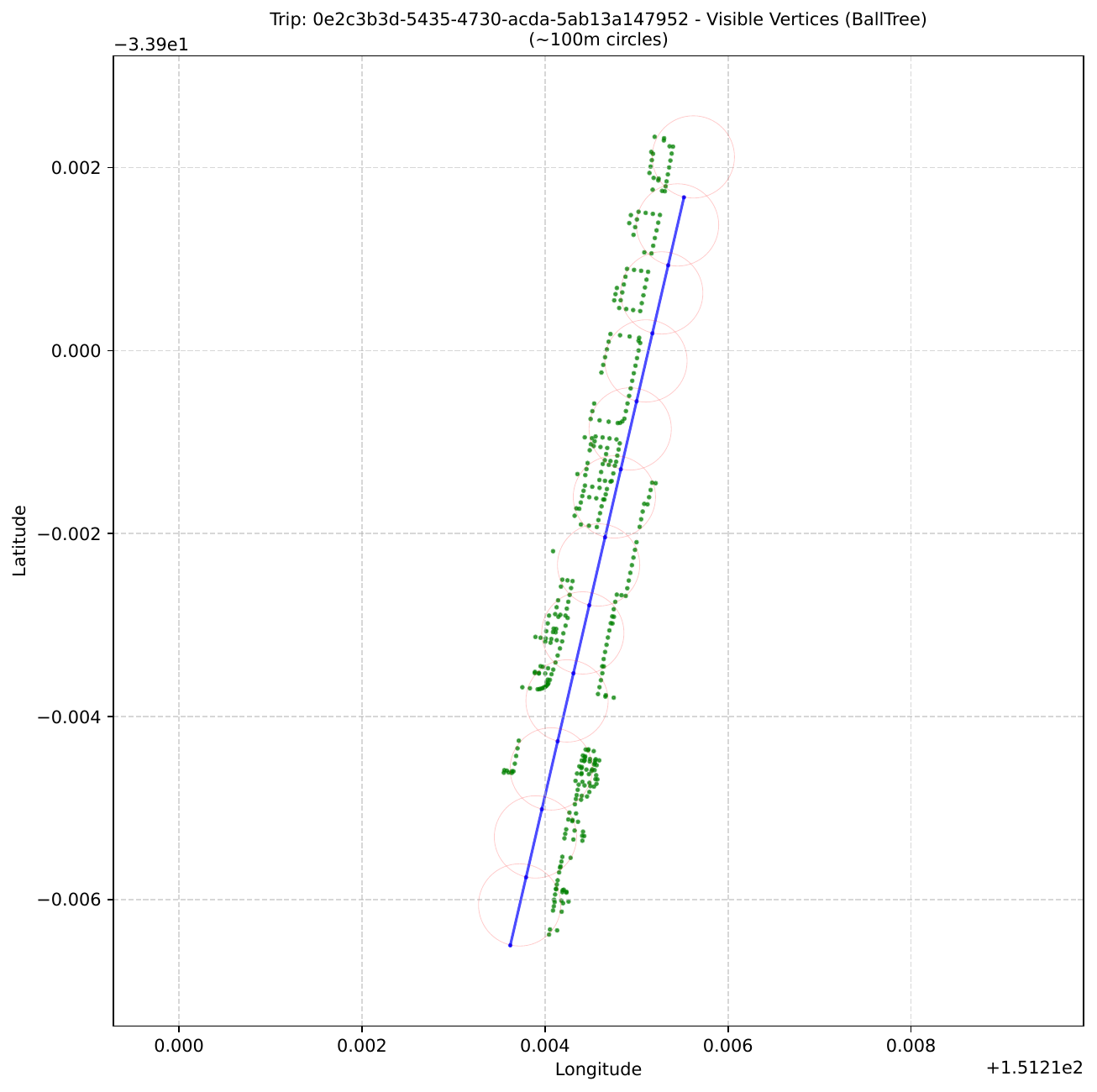}
    \Description{A latitude-longitude scatter of one trip: a straight blue path running lower-left to upper-right, a chain of overlapping pale-red 100-metre viewing circles along it, and green building-vertex dots tracing the frontages on both sides.}
    \caption{Example interpolated trajectory trace and associated viewing circles for a single trip within the Waterloo study area. Blue line indicates vehicle path (48 points), red circles represent the 100m diameter forward viewing areas used for visibility analysis.}
    \label{fig:single_trip_example}
\end{figure*}

Figure~\ref{fig:single_trip_example} illustrates the output of the trajectory processing and the definition of viewing areas used in the visibility modeling for a single vehicle trip (48 interpolated points) within the study area. The blue line represents the vehicle's path derived from the 5-second interpolation, while the red circles depict the 100-meter diameter viewing areas projected 50 meters ahead along the vehicle's bearing at each point. These circles define the regions within which the *densified building points* were queried for visibility using the BallTree index. This visualization helps understand the spatial coverage of the visibility assessment along an individual route.

\subsection{Aggregated Visibility Distribution} 

\begin{figure}[htbp]
    \centering
    \includegraphics[width=0.75\linewidth]{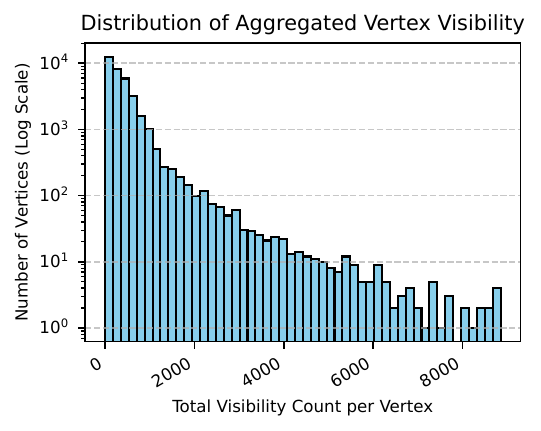}
    \Description{A histogram of visibility count per vertex with a logarithmic vertical axis. It is strongly right-skewed: tallest bars near ten thousand vertices at low counts, falling rapidly into a long thin tail reaching counts around nine thousand.}
    \caption{Frequency distribution of aggregated visibility counts per building vertex across all processed trajectories. The y-axis (Number of Vertices) is plotted on a logarithmic scale.}
    \label{fig:visibility_histogram}
\end{figure}

The aggregated visibility counts across all processed trajectories exhibit a highly skewed distribution, as shown in Figure \ref{fig:visibility_histogram}. This histogram plots the total number of times each unique building vertex was captured within a viewing circle. The x-axis represents the aggregated visibility count, while the y-axis shows the number of vertices falling into each count bin, presented on a logarithmic scale to accommodate the wide range of values. The distribution clearly indicates that a vast majority of building vertices have very low visibility counts, while a small number of vertices achieve significantly higher counts, forming a long tail characteristic of phenomena where exposure is concentrated. This skewness suggests that overall visibility is dominated by relatively few locations within the study area.

\begin{figure*}[htbp]
    \centering
    \includegraphics[width=\linewidth]{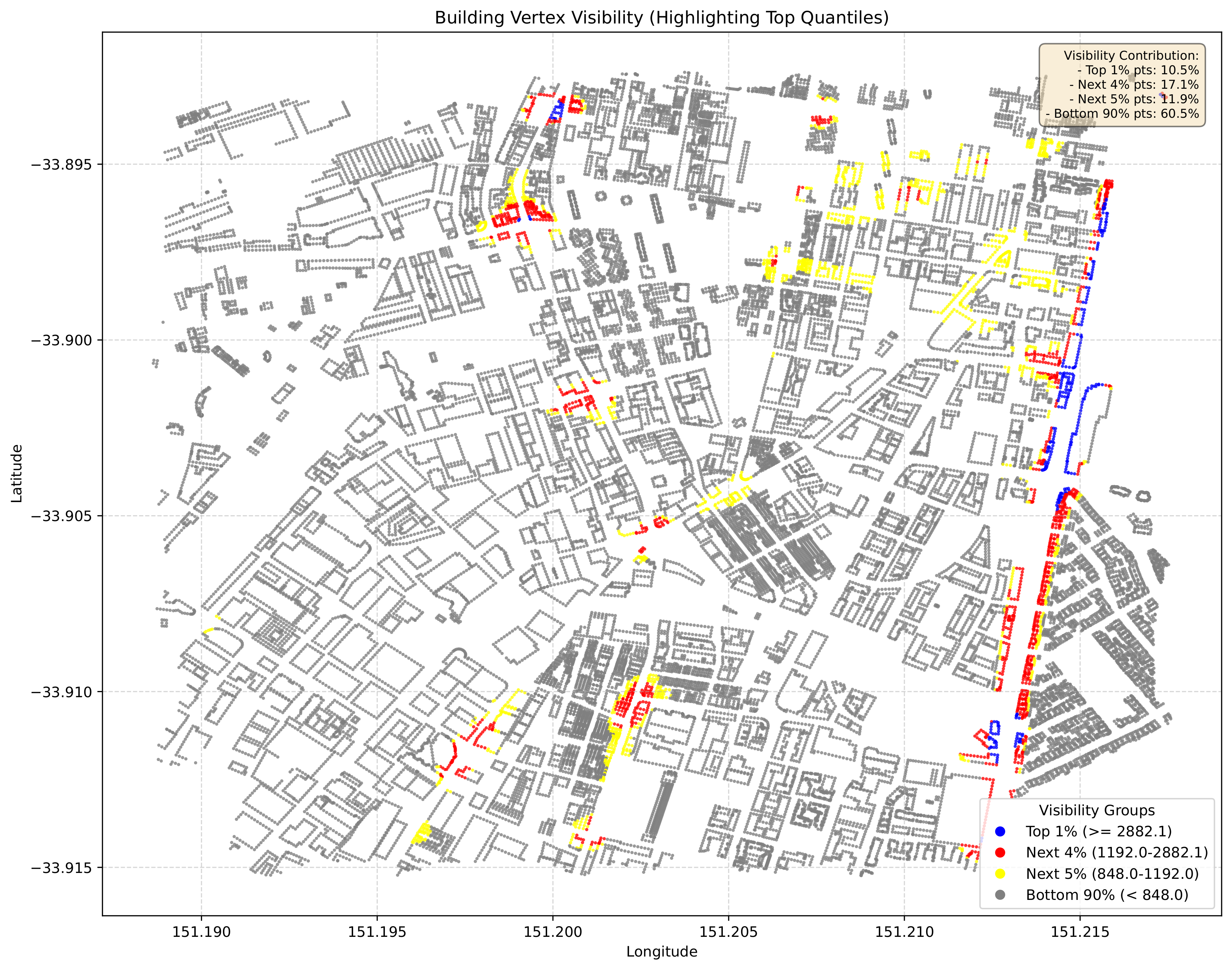}
    \Description{A map of building footprints as black outlines. Bottom-90-percent vertices are grey; higher quantiles are coloured yellow (next 5 percent), red (next 4 percent), and blue (top 1 percent). The coloured vertices concentrate along major thoroughfares and intersections, especially the eastern edge, marking visual hotspots.}
    \caption{Spatial distribution of Building Vertex Visibility, Highlighting Top Quantiles. Different colors represent vertices falling into specific aggregated visibility count percentiles (e.g., Bottom 90\%, 90th-95th, 95th-99th, Top 1\%).}
    \label{fig:top_quantiles_scatter}
\end{figure*}

To further explore the spatial concentration of visibility, Figure \ref{fig:top_quantiles_scatter} visualizes the locations of building vertices color-coded according to their aggregated visibility quantile group. This plot distinctly highlights the vertices falling within the highest visibility percentiles (e.g., top 1\%, 95th-99th percentile) compared to the vast majority of low-visibility points (e.g., bottom 90\%). The visualization confirms the spatial clustering of high-visibility vertices, often located at prominent intersections or along major thoroughfares within the Waterloo area. Furthermore, the plot quantifies the contribution of these top quantile groups to the overall visibility sum, typically demonstrating a Pareto-like principle where a small fraction of vertices accounts for a large percentage of the total visibility events captured across all trajectories (e.g. top 10\% of points gains around 39.5\% of total visibility).

\subsection{Statistical Distribution of Visibility} 

The aggregated visibility counts for the 34,495 unique building vertices with non-zero visibility were fitted to several common probability distributions. The goal was to characterize the overall distribution of visibility across the study area. Table~\ref{tab:dist_fitting_results} summarizes the estimated parameters for each fitted distribution, along with goodness-of-fit statistics including the Kolmogorov-Smirnov (K-S) statistic (D) and the 1-Wasserstein distance. Lower values for D and Wasserstein distance generally indicate a better fit to the empirical data.

\begin{table*}[htbp]
\small
    \centering
    \caption{Distribution Fitting Results for Aggregated Vertex Visibility Counts (N=34,495). Parameters shown are shape (s, a, c), location (loc), and scale. K-S p-values were effectively zero ($< 10^{-7}$) for all fits shown.}
    \label{tab:dist_fitting_results}

    \sisetup{round-mode=places, round-precision=4, table-align-text-post=false}
    \begin{tabular}{l S[table-format=1.4]
                      S[table-format=1.4] 
                      S[table-format=-2.4] 
                      S[table-format=3.4] 
                      S[table-format=1.4] 
                      S[table-format=1.4] 
                      S[table-format=1.4]} 
        \toprule
        \textbf{Distribution} & {\textbf{K-S Stat (D)}} & {\textbf{Wasserstein Dist.}} & {\textbf{loc}} & {\textbf{scale}} & {\textbf{a}} & {\textbf{c}} & {\textbf{s}} \\
        \midrule
        Log-Normal    & 0.0489 & 40.4802 & -23.5784 & 263.5925 & {--} & {--} & 1.0332 \\
        Gamma         & 0.9936 & 413.0533 & 1.0000  & 2.6506   & 0.0029 & {--} & {--} \\
        Exponential   & 0.0493 & 62.1045 & 1.0000  & 413.0612 & {--} & {--} & {--} \\
        Weibull (min) & 0.1324 & 71.2408 & 1.0000  & 299.4076 & {--} & 0.7558 & {--} \\
        Normal        & 0.2358 & 259.2370 & 414.0612 & 573.7056 & {--} & {--} & {--} \\
        Inverse Gamma & 0.0482 & 55.1729 & -108.3701 & 670.4057 & 2.2225 & {--} & {--} \\
        Gumbel (R)    & 0.0858 & 96.7158 & 234.0655 & 259.4164 & {--} & {--} & {--} \\
        \bottomrule
    \end{tabular}
    \caption*{\footnotesize K-S: Kolmogorov-Smirnov statistic. Wasserstein Dist.: 1-Wasserstein distance. Lower values indicate better fit. Parameters a, c, s are shape parameters specific to certain distributions. `--` indicates parameter not applicable for the distribution.}
\end{table*}

Based on the goodness-of-fit statistics, particularly the K-S statistic and the Wasserstein distance, the Log-Normal distribution provides the best fit to the empirical visibility data among the tested distributions, followed by the Inverse Gamma and Exponential distributions. This confirms the highly skewed, long-tailed nature of vertex visibility observed in the histogram (Figure~\ref{fig:visibility_histogram}). The poor fit of the Normal distribution further emphasizes this non-Gaussian characteristic.

\section{Shape similarity}

Table~\ref{tab:pearson_correlation} displays the Pearson correlation coefficient ($r$) for visibility count variables for the study area across three shapes: `Circle`, `Sector`, and `Triangle`. The most important takeaway is that all three variables are strongly and positively correlated with one another. All vertex visibility count correlation coefficients are above 0.92.

\begin{table}[htbp]
\small
\centering
\caption{Pearson Correlation Matrix for vertex visibility}
\label{tab:pearson_correlation}
\begin{tabular}{lccc}
\toprule
& \textbf{Circle} & \textbf{Sector} & \textbf{Triangle} \\
\midrule
\textbf{Circle}    & 1.0000          & 0.9718          & 0.9248            \\
\textbf{Sector}    & 0.9718          & 1.0000          & 0.9464            \\
\textbf{Triangle}  & 0.9248          & 0.9464          & 1.0000            \\
\bottomrule
\end{tabular}
\end{table}

\begin{figure*}[htbp]
    \centering 

    \begin{subfigure}[b]{0.32\textwidth}
        \includegraphics[width=\linewidth]{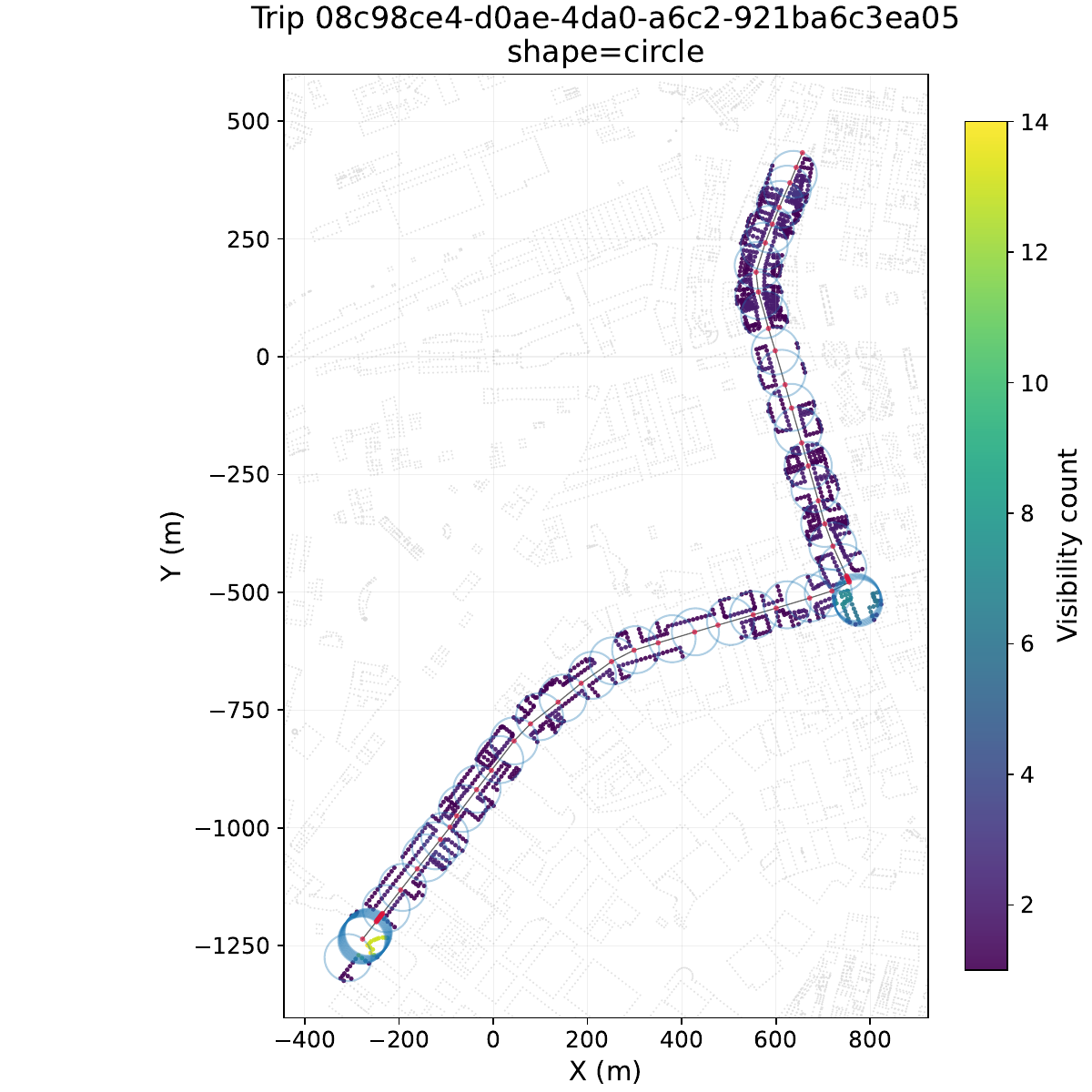}
        \Description{A circular point-of-view trip in X-Y metres: a path running lower-left to upper-right, with building vertices coloured by visibility count on a green scale from zero to about fourteen, densest mid-route and at the upper end.}
        \caption{Circle POV}
        \label{fig:circle_trip}
    \end{subfigure}
    \hfill 
    %
    \begin{subfigure}[b]{0.32\textwidth}
        \includegraphics[width=\linewidth]{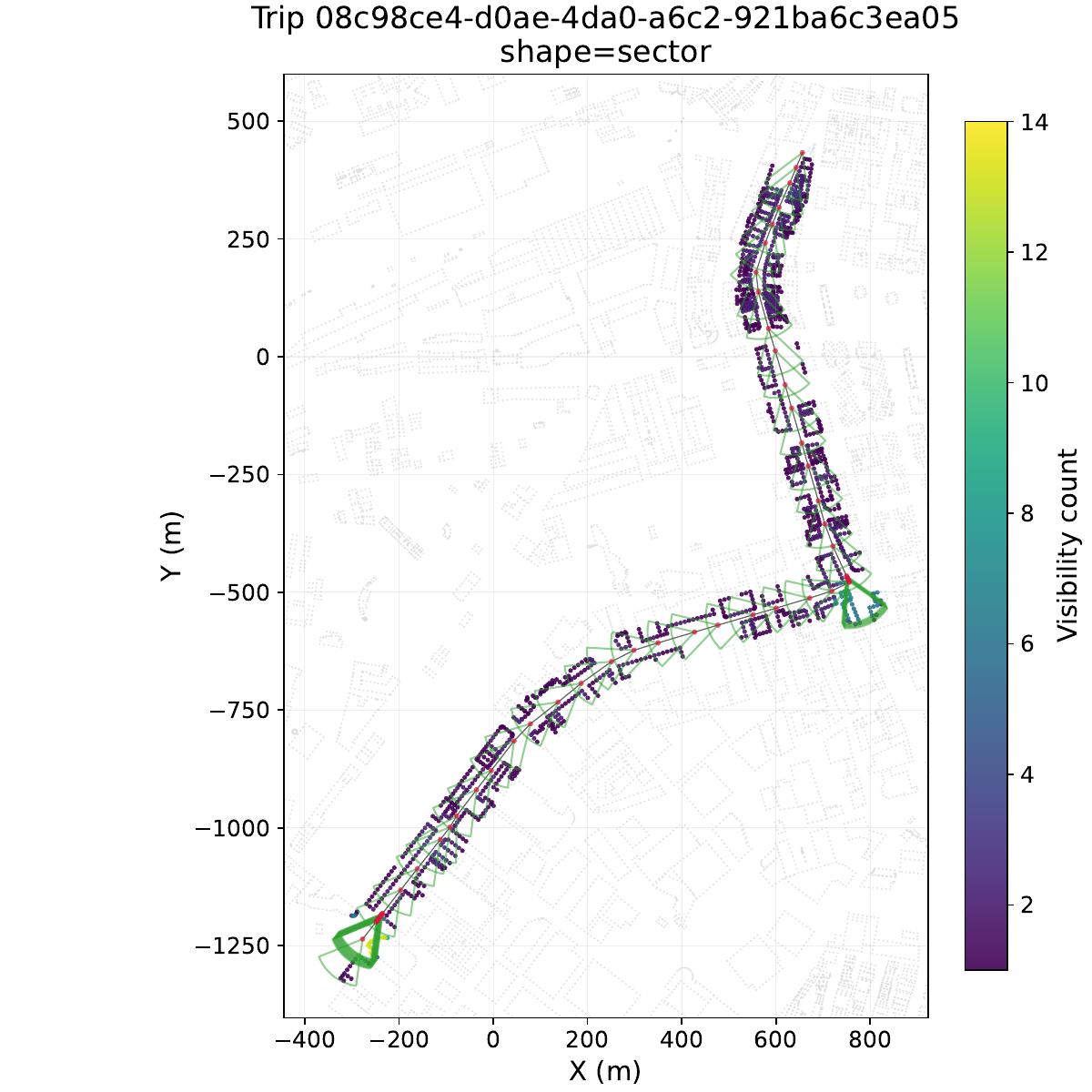}
        \Description{A sector-shaped point-of-view of the same trip; the path and green-scaled visibility-count vertices look almost indistinguishable from the circular panel.}
        \caption{Sector POV}
        \label{fig:sector_trip}
    \end{subfigure}
    \hfill 
    %
    \begin{subfigure}[b]{0.32\textwidth}
        \includegraphics[width=\linewidth]{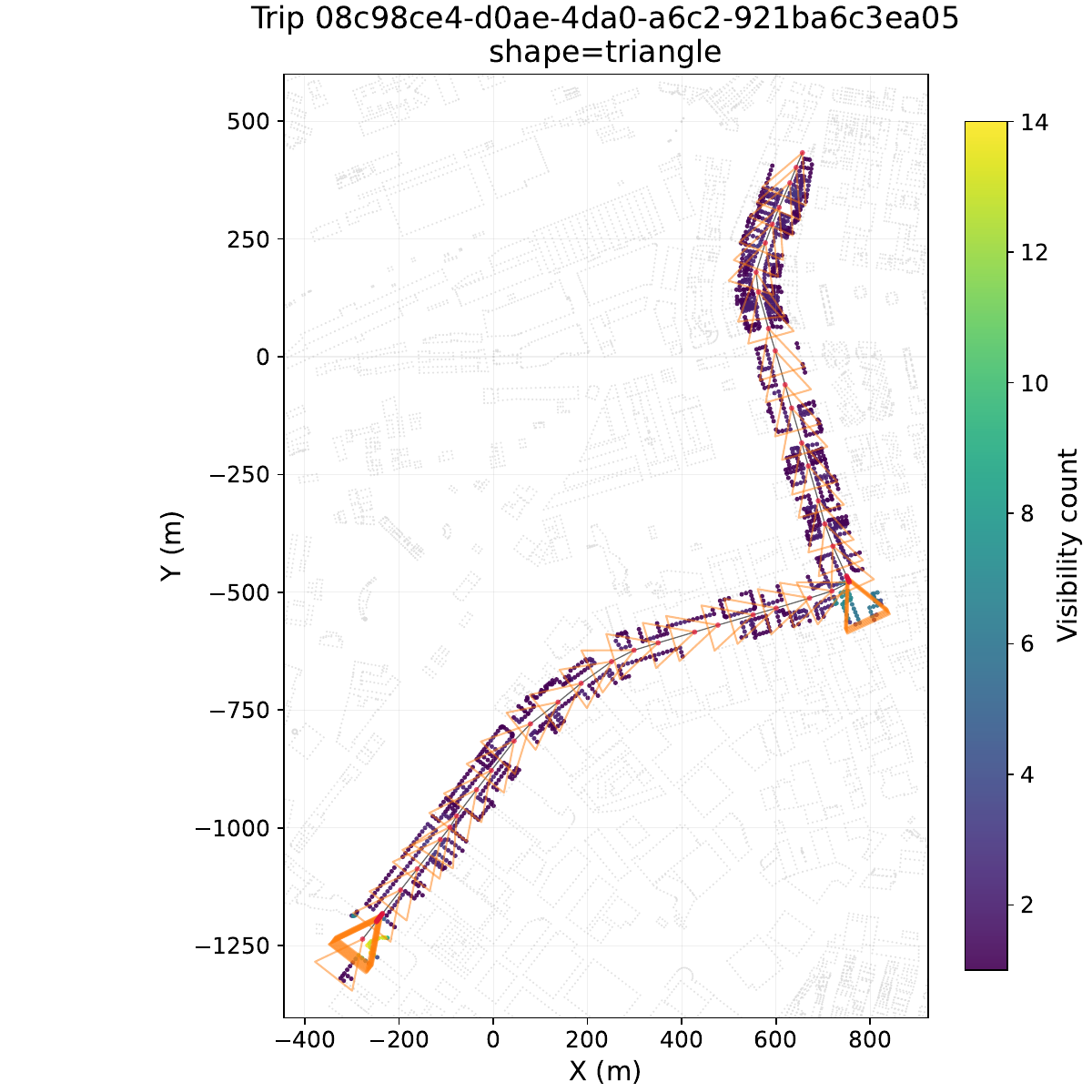}
        \Description{A triangular point-of-view of the same trip; the path and green-scaled visibility-count vertices closely match the circle and sector panels.}
        \caption{Triangle POV}
        \label{fig:triangle_trip}
    \end{subfigure}

    \caption{Comparison of vertex visibility for a sample trajectory using three different Point of View (POV) shapes. The driver's path is shown in black, with sample POV shapes overlaid. Vertices are colored by the number of times they were visible.}
    \label{fig:trip_comparison}
\end{figure*}

To verify this effect, we conducted a visibility analysis along a sample vehicle trajectory using three distinct POV models: an forward-facing circle, a forward-facing triangular cone, and  sector. The resulting visibility maps, which highlight the building vertices "seen" by each model, are compared on the same trajectory in Figure~\ref{fig:trip_comparison}. This comparison visually demonstrates similarity of POV shapes and visibility counts distribution.

\section{Performance Considerations}

\begin{figure*}[h]
  \centering
  \Description{Two log-scale bar charts, index versus brute force across circle, triangle, and sector shapes. Throughput (trajectories per minute) is far higher for index (circle 6044, sector 3762, triangle 3441) than brute force (35, 36.2, 3.8). Latency (seconds per trajectory) is 0.010 to 0.017 for index versus 1.7 to 15.8 for brute force; circle index is fastest.}
  \includegraphics[width=\textwidth]{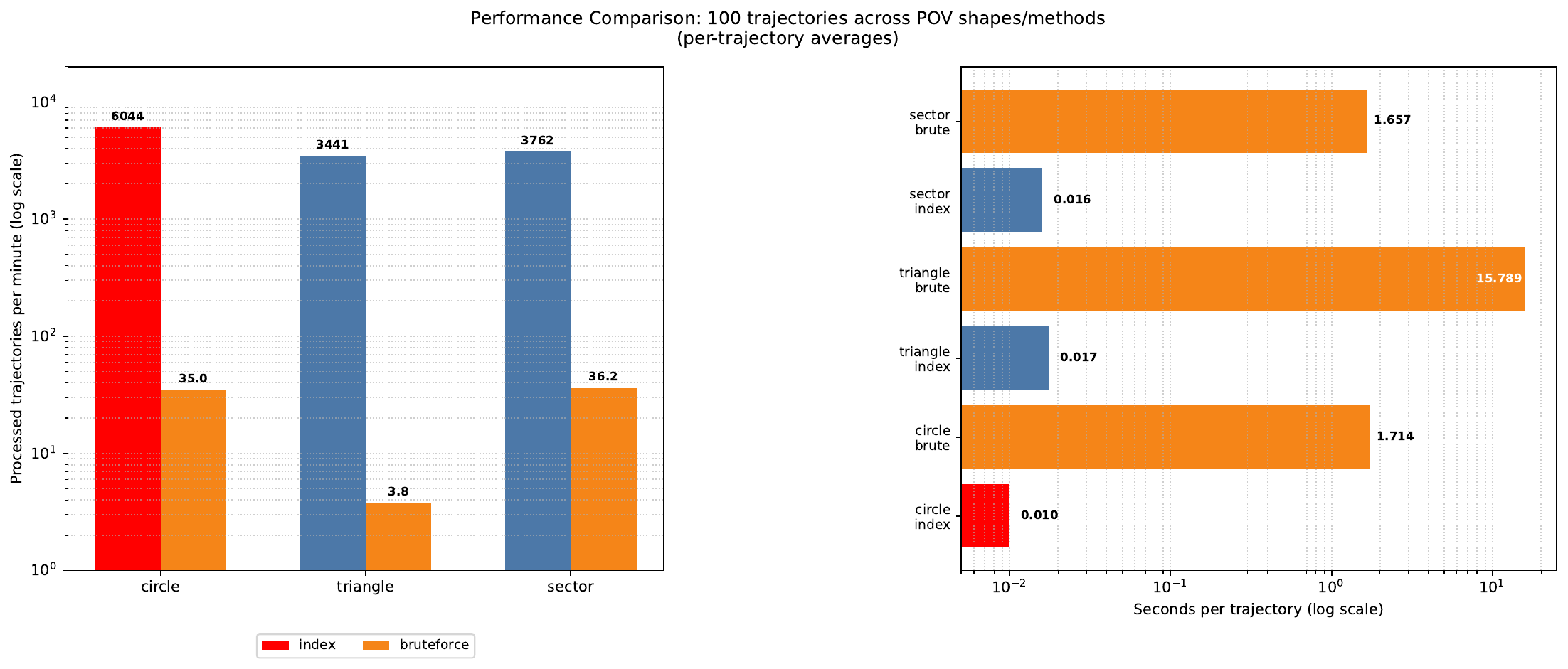}
  \caption{Processing performance (left, trajectories per minute; higher is better) and latency (right, seconds per trajectory; lower is better) for first 100 trips across POV configurations.}
  \label{fig:performance_comparison}
\end{figure*}

The visibility pipeline was benchmarked on the first one hundred trips recorded in the study area, using four point‑of‑view (POV) configurations (see Figure \ref{fig:performance_comparison}). 
All experiments were executed on a single core of an AMD Ryzen 7600 processor with 32 GB of RAM, and with scikit‑learn 1.72.0 for spatial indexing. The circle shape using spatial indexing via balltree method is the fastest, outperforms all other shapes by factor of 1.5-2x even with the same indexing methods used for points filtering. The bruteforce implementation in Python (all building vertexes gets a separate check without filtering) is multiple orders of magnitude slower when not using indexing method. The high processing speed makes it possible to perform visibility analysis at large scale (e.g. city or comparatively study cities, not just single suburb) within reasonable time.

\section{Discussion}
\label{sec:discussion}

The findings of this research provide an overview of the nature of urban visibility from a perspective of vehicular trajectories. 
The primary observation  that building vertex visibility is highly concentrated in specific ``visual hotspots''  has direct implications for Out-of-Home (OOH) advertising and urban planning. It suggests that a large proportion of potential visual exposure originates from a relatively small number of geographical points, corresponding to buildings at major intersections or prominent corners. This contrasts with planning based solely on average traffic volume because it does not account for driving patterns and visibility considerations (direction and area), which might overlook these high-impact locations or overvalue less visually prominent ones.

Furthermore, the strong adherence of the aggregated visibility counts to a Log-Normal distribution is noteworthy. This statistical characteristic, often found in phenomena involving multiplicative growth or compounding factors, implies that the factors influencing high visibility (e.g., road geometry, traffic speed variations, traffic flow patterns, building prominence, duration of view) likely interact in a way that creates extreme values. This quantitative characterization provides a basis for more sophisticated analysis, such as predictive modeling or differential valuation of potential advertising sites based on their position in the visibility distribution.

The methodology presented, utilizing connected vehicle data, OpenStreetMap features, and BallTree spatial indexing, demonstrates a computationally efficient approach to generating these granular visibility metrics at scale. The resulting quantitative visibility data and hotspot identification can support decisions regarding the placement of OOH assets like billboards and digital screens for improving campaign reach and effectiveness. Similarly, urban designers and planners can use this information to identify locations requiring high public visibility for signage, street furniture, or public information displays.

The current visibility model implements a simplified 2D forward-looking circle, which represents the perception area of a vehicle driver. This geometric abstraction does not account for 3D occlusions caused by other vehicles or vehicle interior, varying building heights, or roadside vegetation, nor does it incorporate factors central to visual perception like specific viewing angles, the dynamics of driver attention, or the characteristics of the advertisement itself.

The connected vehicle data, while providing spatial coverage within the study area (Waterloo, Sydney), represents only a sample of the total traffic volume. Its representativeness across all vehicle types (private cars, commercial vehicles, public transport) warrants further investigation to understand potential biases in the visibility patterns observed.

\section{Conclusion}
\label{sec:conclusion_final}

This research addressed the need for more accurate, dynamic, and data-driven methods for assessing the visual exposure of building features in complex urban environments. We developed and demonstrated a scalable computational framework that utilizes large-scale connected vehicle trajectory data (Compass IoT) and open geospatial building data (OpenStreetMap) for the Waterloo area in Sydney. By applying time-based trajectory interpolation, defining a forward-projected viewing area, and utilizing efficient BallTree spatial indexing, our methodology quantifies the cumulative visibility counts for thousands of building vertices based on observed, real-world vehicle movement patterns.

The analysis revealed that building vertex visibility is highly concentrated in specific 'visual hotspots' and that the overall distribution of visibility strongly conforms to a Log-Normal model. The primary contribution of this work is the provision of a computationally efficient method for generating granular, objective visibility metrics, offering a significant improvement over traditional static or volume-based assessments. These findings are relevant for enhancing urban design, informing location-based analytics, and guiding the placement of street-level infrastructure such as signage and OOH advertising assets. While acknowledging the limitations of the current visibility model and data scope, this study highlights the substantial potential of combining large-scale mobility data with advanced spatial analysis techniques to better understand and quantify visual exposure patterns within dynamic city environments. 

A notable limitation is the model's inability to account for 3D occlusions. For example, at the intersection of Bourke Street and Ebsworth Street (coordinates: -33.905499, 151.203835), a tall commercial building blocks visibility to lower buildings despite high visibility counts. Future work will incorporate digital elevation models to address this constraint.

Further research may also explore the temporal dynamics of visibility  how patterns change by time of day, day of week, or due to traffic conditions  presents another promising avenue for refining visibility estimates. Crucially, validating the derived visibility metrics against empirical data, such as human eye-tracking studies conducted in real or simulated environments, or correlating metrics with downstream outcomes of interest, remains an essential step to fully establish the practical predictive power and utility of this quantitative visibility framework. Future work can also be directed towards a bigger scale both in space and time, may focus on analysis of real-time data or calculating change in visibility patterns before, during and after major events.

\begin{acks}
We thank Compass IoT for the data and support provided for this study. This work has been funded by the UTS Jenny Edwards Fellowship awarded in 2025 to Assoc. Prof. Adriana-Simona Mihaita for conducting research on connected vehicles. 

Anthropic Claude was used to format the manuscript according to the ACM template.
\end{acks}

\bibliographystyle{ACM-Reference-Format}
\bibliography{references}       

@article{Fong2013Identifying,
author = {Fong, Simon and Cho, Kyungeun and Ip, Weng and Liu, Elaine},
year = {2013},
month = {07},
pages = {},
title = {Identifying Optimal Spatial Groups for Maximum Coverage in Ubiquitous Sensor Network by Using Clustering Algorithms},
volume = {2013},
journal = {International Journal of Distributed Sensor Networks},
doi = {10.1155/2013/763027}
}

@article{Chmielewski2017,
  author       = {Chmielewski, Szymon and Tompalski, Piotr},
  title        = {Estimating outdoor advertising media visibility with voxel-based approach},
  journal      = {Applied Geography},
  volume       = {87},
  pages        = {1--13},
  year         = {2017},
  issn         = {0143-6228},
  doi          = {10.1016/j.apgeog.2017.07.007}
}

@article{Madlenak2023,
  author       = {Madle\v{n}\'{a}k, Radovan and Chinorack\'{y}, Roman and Stalma\v{s}ekov\'{a}, Nat\'{a}lia and Madle\v{n}\'{a}kov\'{a}, Lucia},
  title        = {Investigating the Effect of Outdoor Advertising on Consumer Decisions: An Eye-Tracking and {A/B} Testing Study of Car Drivers' Perception},
  journal      = {Applied Sciences},
  volume       = {13},
  number       = {11},
  pages        = {6808},
  year         = {2023},
  doi          = {10.3390/app13116808}
}

@book{Clow2023Integrated,
  author       = {Clow, Kenneth E. and Baack, Donald E.},
  title        = {Integrated Advertising, Promotion, and Marketing Communications, Global Edition},
  edition      = {7th},
  publisher    = {Pearson},
  year         = {2016}
}

@article{SmartAdP,
  author    = {Liu, Dongyu and Weng, Di and Li, Yuhong and Bao, Jie and Zheng, Yu and Qu, Huamin and Wu, Yingcai},
  title     = {{SmartAdP}: Visual Analytics of Large-scale Taxi Trajectories for Selecting Billboard Locations},
  journal   = {IEEE Transactions on Visualization and Computer Graphics},
  volume    = {23},
  number    = {1},
  pages     = {1--10},
  year      = {2017},
  doi       = {10.1109/TVCG.2016.2598432}
}

@ARTICLE{liu2024modeling,
  author={Farah, Haneen and Postigo, Ivan and Reddy, Nagarjun and Dong, Yongqi and Rydergren, Clas and Raju, Narayana and Olstam, Johan},
  journal={IEEE Transactions on Intelligent Transportation Systems}, 
  title={Modeling Automated Driving in Microscopic Traffic Simulations for Traffic Performance Evaluations: Aspects to Consider and State of the Practice}, 
  year={2023},
  volume={24},
  number={6},
  pages={6558-6574},
  doi={10.1109/TITS.2022.3200176}}

@phdthesis{DevelopmentEvaluationSimulation,
author = {Safaei Matin, Ali},
title = {Development and evaluation of simulation models for assessing the impacts of connected and automated vehicles},
year = {2024},
school = {Swinburne University of Technology},
type = {Ph.D. Dissertation},
url = {https://figshare.swinburne.edu.au/articles/thesis/Development_and_evaluation_of_simulation_models_for_assessing_the_impacts_of_connected_and_automated_vehicles/28087664},
doi = {10.25916/sut.28087664.v1}
}

@inproceedings{AnAgentBasedModel,
author = {Huynh, Nam and Cao, Vu Lam and Wickramasuriya, Rohan and Berryman, Matthew and Perez, Pascal and Barth\'{e}lemy, Johan},
year = {2014},
title = {An Agent Based Model for the Simulation of Road Traffic and Transport Demand in a Sydney Metropolitan Area},
booktitle = {Proceedings of the 8th International Workshop on Agents in Traffic and Transportation (ATT 2014)},
address = {Paris, France},
pages = {1--7},
doi = {10.13140/2.1.4023.2961}
}

@article{AgentbasedSimulationPedestrian,
title = {Agent-based simulation for pedestrian evacuation: A systematic literature review},
journal = {International Journal of Disaster Risk Reduction},
volume = {111},
pages = {104705},
year = {2024},
issn = {2212-4209},
doi = {https://doi.org/10.1016/j.ijdrr.2024.104705},
author = {Gayani P.D.P. Senanayake and Minh Kieu and Yang Zou and Kim Dirks}
}

@article{ModellingImpactTransit,
author = {Ma{\l}ecki, Krzysztof and Jankowski, Jaros{\l}aw and Szkwarkowski, Mateusz},
year = {2019},
month = {03},
pages = {428},
title = {Modelling the Impact of Transit Media on Information Spreading in an Urban Space Using Cellular Automata},
volume = {11},
journal = {Symmetry},
doi = {10.3390/sym11030428}
}

@book{HandbookMobilityDataMining,
title = {Handbook of Mobility Data Mining, Volume 3: Mobility Data-Driven Applications},
editor = {Zhang, Haoran},
publisher = {Elsevier},
year = {2023},
isbn = {978-0-323-95892-9}
}

@phdthesis{StudyMapMatching,
  author      = {Chao, Pingfu},
  title       = {A study on map-matching and map inference problems},
  school      = {School of Information Technology and Electrical Engineering},
  institution = {The University of Queensland},
  year        = {2020},
  month       = aug,
  doi         = {10.14264/uql.2020.1009},
  url         = {https://doi.org/10.14264/uql.2020.1009},
  type        = {PhD Thesis},
  language    = {eng}
}

@article{SolvingDataSparsity,
  author    = {Xue, Andy Yuan and Qi, Jianzhong and Xie, Xing and Zhang, Rui and Huang, Jin and Li, Yuan},
  title     = {Solving the data sparsity problem in destination prediction},
  journal   = {The VLDB Journal},
  volume    = {24},
  number    = {2},
  pages     = {219--243},
  year      = {2015},
  month     = apr,
  issn      = {0949-877X},
  doi       = {10.1007/s00778-014-0369-7},
  url       = {https://doi.org/10.1007/s00778-014-0369-7}
}

@article{TargetedAdvertisingThesis,
  author    = {Faroqi, Hamed and Mesbah, Mahmoud and Kim, Jiwon and Khodaii, Ali},
  title     = {{Targeted Advertising in the Public Transit Network Using Smart Card Data}},
  journal   = {Networks and Spatial Economics},
  volume    = {22},
  number    = {1},
  pages     = {97--124},
  year      = {2022},
  month     = mar,
  issn      = {1572-9427},
  doi       = {10.1007/s11067-022-09558-9},
  url       = {https://doi.org/10.1007/s11067-022-09558-9}
}

@ARTICLE{ScalableFrameworkTrajectoryPrediction,
  author={Rathore, Punit and Kumar, Dheeraj and Rajasegarar, Sutharshan and Palaniswami, Marimuthu and Bezdek, James C.},
  journal={IEEE Transactions on Intelligent Transportation Systems}, 
  title={A Scalable Framework for Trajectory Prediction}, 
  year={2019},
  volume={20},
  number={10},
  pages={3860-3874},
  doi={10.1109/TITS.2019.2899179}}

@article{xiao2022generalized,
  title={A generalized trajectories-based evaluation approach for pedestrian evacuation models},
  author={Xiao, Yao and Xu, Jun and Chraibi, Mohcine and Zhang, Jun and Gou, Chao},
  journal={Safety Science},
  volume={147},
  pages={105574},
  year={2022},
  publisher={Elsevier},
  doi={10.1016/j.ssci.2021.105574}
}

@phdthesis{gschwend2015relating,
  title={Relating movement to geographic context: effects of preprocessing, relation methods and scale},
  author={Gschwend, Christian},
  year={2015},
  school={University of Zurich}
}

@article{KEROUANTON2024100734,
title = {Eye-catching or breath-catching: Role and landscape attributes of pauses differs among hikers' profile when rambling in a French mountainous area},
journal = {Journal of Outdoor Recreation and Tourism},
volume = {46},
pages = {100734},
year = {2024},
issn = {2213-0780},
doi = {https://doi.org/10.1016/j.jort.2024.100734},
url = {https://www.sciencedirect.com/science/article/pii/S2213078024000021},
author = {Colin Kerouanton and Laurence Jolivet and Cl\'{e}mence Perrin-Malterre and Anne Loison}
}

\end{document}